\documentclass[11pt,a4paper]{article}
\usepackage{epsfig,amsmath,amstext,amssymb,bm,comment,color,graphicx,subfigure,epstopdf}

\newcommand{\smooth}{\mathfrak{R}}
\newcommand{\sm}{\theta}

\begin{document}

\title{\bf Chiral Superfluidity for QCD}
\author{T.~Kalaydzhyan\footnote{{\bf e-mail}: tigran.kalaydzhyan@stonybrook.edu}
\\
\small{\em Department of Physics and Astronomy, Stony Brook University,} \\
\small{\em Stony Brook, NY 11794-3800, U.S.A.}\\
}
\date{}
\maketitle

\begin{abstract}
We argue that the strongly coupled quark-gluon plasma formed at LHC and RHIC can be considered as a chiral superfluid. The ``normal'' component of the fluid is the thermalized matter in common sense, while the ``superfluid'' part consists of long wavelength (chiral) fermionic states moving independently. We use the bosonization procedure with a finite cut-off and obtain a dynamical axion-like field out of the chiral fermionic modes. Then we use relativistic hydrodynamics for macroscopic description of the effective theory obtained after the bosonization. Finally, solving the hydrodynamic equations in gradient expansion, we find that in the presence of external electromagnetic fields or rotation the motion of the ``superfluid'' component gives rise to the chiral magnetic, chiral vortical,  chiral electric and dipole wave effects. Latter two effects are specific for a two-component fluid, which provides us with crucial experimental tests of the model.
\end{abstract}
\vspace{.7cm}

\section*{Introduction}

The non-trivial structure of the QCD vacuum attracted much attention in light of recent heavy-ion experiments performed at RHIC and LHC. These experiments make it possible to study the strongly-coupled quark-gluon plasma (sQGP) in hadronic scale magnetic fields \cite{Tuchin:2013ie}. The non-trivial gluonic configurations may induce an imbalance between densities of left- and right-handed light quarks (chirality). As a strong magnetic field is applied to the system, the imbalance can give rise to a net electric current in sQGP along the magnetic field (chiral magnetic effect \cite{Kharzeev:2007jp, CME}). So far it was difficult to build a first-principles theory, describing this and similar effects, since the physics of sQGP is essentially nonperturbative. However, there is a need in such a theory, because constant axial chemical potentials, introduced by hand, break unitarity \cite{Adam:2001ma} and lead to various consistency and stability problems \cite{Khaidukov:2013sja, Avdoshkin:2014gpa}. At the same time, without the knowledge of all possible anomalous effects of a similar kind, one will face difficulties in the experimental searches for each of them. Fortunately, it seems that such a theory can be established (and it is sketched below), because QCD contains a long-wave axion-like degree of freedom, which can play a role of carrier for the chirality. Indeed, one can consider QCD coupled to QED with an auxiliary gauged $U_A(1)$, and bosonize quarks with Dirac eigenvalues smaller than some fixed scale $\Lambda$. Since there is no gauge $U_A(1)$ in nature, one can choose a pure gauge form of the external axial vector field and, as a result of the procedure, one obtains the following effective Euclidean Lagrangian \cite{kalaydzhyan, Damgaard:1993sx},
	        \begin{align}
		{\cal L}_E &= \frac{1}{4}G^{a\mu\nu}G_{\mu\nu}^{a} + \frac{1}{4}F^{\mu\nu}F_{\mu\nu} -j^{\mu} A_\mu - g j^{a\,\mu} G^a_\mu \nonumber\\
 		& + \displaystyle\frac{\Lambda^2 N_c}{4\pi^2} \partial^\mu\theta\,\partial_\mu\theta\ +\frac{g^{2}}{16\pi^{2}}\theta G^{a\mu\nu}\tilde{G}{}_{\mu\nu}^{a} + \frac{N_c}{8\pi^{2}}\theta F^{\mu\nu}\tilde{F}{}_{\mu\nu}\nonumber\\
 		& +\displaystyle\frac{N_c}{24\pi^2}\theta \Box^2 \theta - \frac{N_c}{12\pi^2} \left( \partial^\mu\theta\partial_\mu\theta\right)^2 + \mathcal{O}\left(1/\Lambda^2\right)\,,\label{lagrangian}
		\end{align}
where the axion-like field $\theta$ originates from the fermionic IR modes, while source currents can be formed by the UV modes.
The physical value for the cut-off $\Lambda$ is different in different regimes and can be found from the dependence of the chirality on the axial chemical potential \cite{kalaydzhyan}: $ \Lambda = \pi\sqrt{2\, T^2/3 + {2\mu^2}/{\pi^2}}$ at high temperatures; $\displaystyle\Lambda = 2\sqrt{|eB|}$ at strong magnetic fields; $\displaystyle\Lambda \simeq 3\, \mathrm{GeV}$ at small temperatures and weak (or absent) magnetic fields. The $\theta$-field can be deduced from QCD even without mentioning its microscopic origins, see \cite{Iatrakis:2014dka} for a non-dynamical case, where $\theta(\vec x, t)$ was introduced formally as a source of the topological charge density density and generates similar currents. One more interesting case, when anomalous effects are reproduced, is an axionic field, coupled to the topological charge density, but propagating in a bulk of a holographic model \cite{Jimenez-Alba:2014iia}.  As a last remark, it is important that the $\theta$-field is \textit{not} a ghost, which can be seen from the sign of its kinetic term in (\ref{lagrangian}) and from the analysis of Ref.~\cite{Damgaard:1993sx}.

\section*{Chiral superfluidity and the chiral magnetic/electric effect}
Chiral superfluidity is a hydrodynamic model of the sQGP in the range of temperatures $T_c < T \lesssim 2\,T_c$ derived from the effective Lagrangian (\ref{lagrangian}) (see \cite{kalaydzhyan} for details).
The essential idea of the model is to consider the sQGP as a two-component fluid with two independent motions: a curl-free motion of the near-zero quark modes (the ``superfluid'' component); and the thermalized medium of dressed quarks (the ``normal'' component). The first one is described by a pseudo-scalar
 field $\theta$, while the second one is represented by a four-velocity $u^\mu$.
Hydrodynamic equations
 can be written in the following way \footnote{here we consider only
 color-singlet currents and neglect the anomaly induced by gluonic fields for simplicity. For the complete treatment see \cite{kalaydzhyan}}
\begin{align}
\partial_{\mu}&T^{\mu\nu} = F^{\nu\lambda} J_{\lambda}\,,\\
\partial_\mu& J^\mu = 0\,,\\
\partial_\mu& J_5^\mu = -\frac{C}{4}  F^{\mu\nu} \tilde F_{\mu\nu}\,,\\
u^\mu& \partial_\mu \theta + \mu_5 = 0\,,\label{Josephson}
\end{align}
where $C$ is the axial anomaly coefficient, e.g. $C = {N_c}/{2\pi^2}$ and the ``axion decay constant'' $f = 2 \Lambda \sqrt{N_c}/\pi$. The constitutive relations can be written in the gradient expansion,
\begin{align}
 T^{\mu\nu} &= \left(\epsilon + P\right)u^\mu u^\nu + P g^{\mu\nu} + f^2 \partial^\mu \theta\partial^\nu \theta+ \tau^{\mu\nu} \,,\label{cons1}\\
 J^\mu & \equiv j^\mu + \Delta j^\mu  = \rho u^\mu + C \tilde{F}^{\mu\kappa} \partial_\kappa \theta + \nu^\mu \,,\label{cons2}\\
 J_5^\mu &=  f^2 \partial^\mu \theta + \nu_5^\mu\,.\label{cons3}
\end{align}
Here $T^{\mu\nu}$ is the energy-momentum tensor of the liquid (total energy-momentum tensor minus the ones of free vector fields), $J^\mu$ is the total electric current, $J_5^\mu$ is the axial current.

The listed equations are similar to the ones of a relativistic superfluid \cite{Son2000}, suggesting the name ``chiral superfluid'' for our model. Eq.~(\ref{Josephson}) is the Josephson equation. One should, of course, distinguish our situation from the conventional superfluidity as, first, there is no spontaneously broken symmetry and, second, the normal component is axially neutral.

Dissipative corrections $\tau^{\mu\nu}$, $\nu^\mu$, $\nu_5^\mu$ are due to finite viscosity, electric resistivity, etc., and do not contain terms proportional to the chiral anomaly coefficient $C$ \cite{kalaydzhyan}.
Additional electric current $\Delta j_\lambda \equiv C \tilde{F}_{\lambda\kappa} \partial^\kappa \theta$ is induced by the ``superfluid'' component and can be represented by a sum of three terms
\begin{align}
\Delta j_\lambda = - C \mu_5 B_\lambda + C \epsilon_{\lambda\alpha\kappa\beta}u^\alpha \partial^\kappa \theta E^\beta - C u_\lambda (\partial \theta \cdot B)\,,\label{phenom}
\end{align}
where the electric and magnetic fields in the ``normal component'' rest frame are defined as
\begin{align}
\label{EB}
 E^\mu = F^{\mu\nu}u_\nu, \qquad B^\mu = \tilde F^{\mu\nu}u_\nu \equiv  \frac{1}{2}\epsilon^{\mu\nu\alpha\beta}u_\nu F_{\alpha\beta}.
\end{align}

The first term in (\ref{phenom}) corresponds to the chiral magnetic effect \cite{Kharzeev:2007jp,CME}, i.e. generation of an electric current along the external magnetic field. The second one is the chiral electric effect \cite{CEE} -- generation of the current transverse to the electric field. The third term we call the chiral dipole effect, which is a generation of electric charge on the inhomogeneities of $\theta$ field. If the profile of $\theta$ is a solitary wave, then this effect will induce a dipole moment on it.
It is worth mentioning that the field $\theta$ itself is a 4D generalization of the chiral magnetic wave \cite{CMW}). In the presence of a strong magnetic field, in becomes exactly the one described in \cite{CMW}, because all higher order terms in (\ref{lagrangian}) disappear in the $\Lambda \propto \sqrt{eB} \rightarrow \infty$ limit, making the bosonization procedure exact, which is a manifestation of the dimensional reduction of the quark dynamics to 2D. 

\section*{Chiral vortical effect}

Finally, let us consider an effect of rotation on the chiral superfluid. The idea will be similar to the one considered in Ref.~\cite{Kalaydzhyan:2014bfa}.
Vorticity (a measure of rotation) of the superfluid component is defined on the superfluid velocity, $u_s^\mu = \partial^\mu \theta / \mu_5$, via
\begin{align}
\omega_\mu = \frac{1}{2} \epsilon_{\mu\nu\alpha\beta} u_s^\nu \partial^\alpha u_s^\beta\,.\label{vorticity}
\end{align}
  In derivation of the Lagrangian (\ref{lagrangian}) in Ref.~\cite{kalaydzhyan}, the $\theta$ field was assumed to be smooth. In this case the superfluid component is irrotational, because of the contraction of the Levi-Civita symbol with two derivatives in (\ref{vorticity}). However, in the presence of rotation, the superfluid component can acquire a nonzero total vorticity through the development of singularities, similar to axionic strings \cite{Callan:1984sa, Kirilin:2012mw},
\begin{equation}
[\partial^\perp_\alpha,\partial^\perp_\beta]\theta=2\pi \delta^{(2)}(\vec{x}_\perp),
\label{eq:defect}
\end{equation}
where $\alpha$ and $\beta$ are in the plane perpendicular to the vortex line (string).
This form of singularity quantizes the vorticity as $\Omega_{quant}= - \pi /\mu_5$ per vortex.\footnote{Additional prefactors in r.h.s. of (\ref{eq:defect}) would not change the final result.} The presence of the delta-function in (\ref{eq:defect}) creates problems, as, e.g., validity of the derivative expansion. Following an elaborated study \cite{Harvey:2000yg}, we resolve them by introducing a smoothing function $\smooth=\smooth(|\vec{x}_\perp|)$, 
\begin{align}
\smooth(|\vec{x}_\perp|\to 0)=-1,\qquad \smooth(|\vec{x}_\perp|\to \infty) = 0\,,
\end{align}
which modifies the Chern-Simons part of the effective action (\ref{lagrangian}),
\begin{align}
\epsilon^{\mu\nu\alpha\beta} \partial_\mu \sm A_\nu F_{\alpha\beta}\, \rightarrow\, \epsilon^{\mu\nu\alpha\beta}(1+\smooth) \partial_\mu \sm A_\nu F_{\alpha\beta}\,.
\end{align}
This takes into account the near-string microscopic physics and introduces a core of the string with a finite thickness. We do not need the exact profile of $\smooth$ for our studies. Varying the effective action with respect to $A_\mu$ we obtain the electric current
\begin{align}
j^\mu = \frac{N_c}{2\pi^2} (1+\smooth) \tilde F^{\mu\nu}\partial_\nu \sm - \frac{N_c}{2\pi}\epsilon^{\mu\nu}A_\nu\tilde\delta^{(2)}(\vec x_\perp)\,,
\end{align}
Where $\tilde\delta^{(2)}$ should be understood as a smoothed delta-function, see \cite{Harvey:2000yg}. Here the first term is identical to (\ref{phenom}) away from the string. The second term is a new current along the string. Let us consider a particular case of rotation around $z$ axis and the presence of (ordinary) chemical potential $\mu$, which can be introduced via $A_\mu = (\mu, \vec 0)$. The current can be rewritten then as
\begin{align}
J^z_{CVE} = -\frac{N_c}{2\pi}\epsilon^{z 0}A_0 \cdot \frac{\Omega^z_{tot}}{\Omega_{quant}} = \frac{N_c}{2\pi}\epsilon^{z 0}A_0 \cdot \frac{\mu_5}{\pi} \Omega^z_{tot} = - C \mu \mu_5 \Omega^z_{tot}\,,
\end{align}
which is the chiral vortical effect in our superfluid, compare to other hydrodynamic results, e.g., \cite{Kirilin:2012mw, Sadofyev:2010pr, Kalaydzhyan:2011vx}. Introduction of $\smooth(|\vec{x}_\perp|)$ separates nicely the bulk currents from the string currents.

Finally, the electric field parallel to the vortex line, if present, will break the gauge-invariance on the string,
\begin{align}
\partial_\mu j^\mu = -\frac{N_c}{4 \pi}\epsilon^{\mu\nu}F_{\mu\nu} \delta^{(2)}(\vec x_\perp)\,,
\end{align}
but not in the full theory due to the induced anomaly inflow, see e.g. \cite{Callan:1984sa, Harvey:2000yg}. Physically, the electric field will induce bulk electric currents, going radially toward the vortex lines, and then flowing into and along them.

\section*{Toward the microscopic picture}

\begin{figure}[t]
     \centering
     \subfigure[\label{below}]
     {\includegraphics[width=4cm]{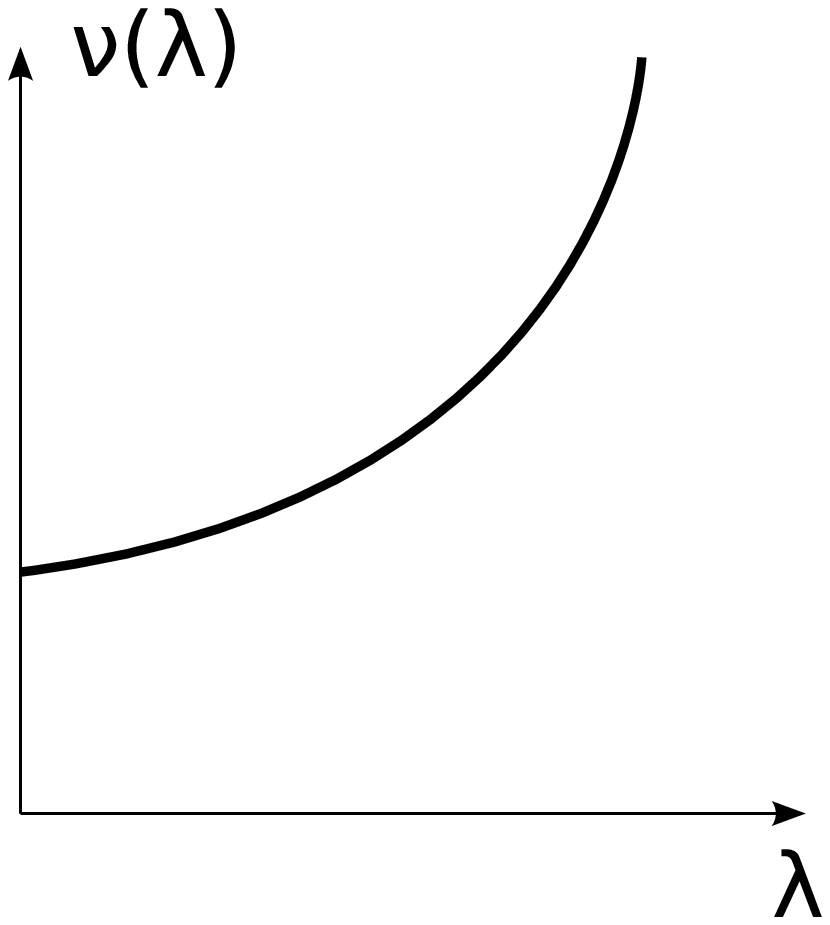}}\hspace{1cm}
     \centering
     \subfigure[\label{above}]
     {\includegraphics[width=4cm]{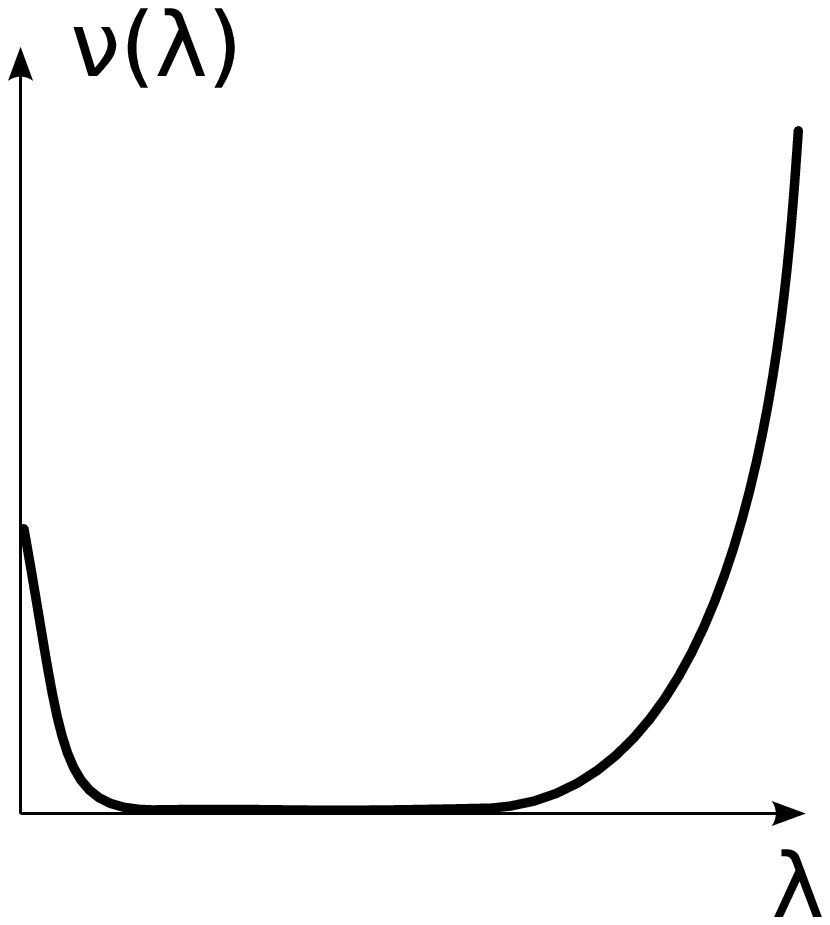}}\hspace{1cm}
     \centering
     \subfigure[\label{high}]
     {\includegraphics[width=4cm]{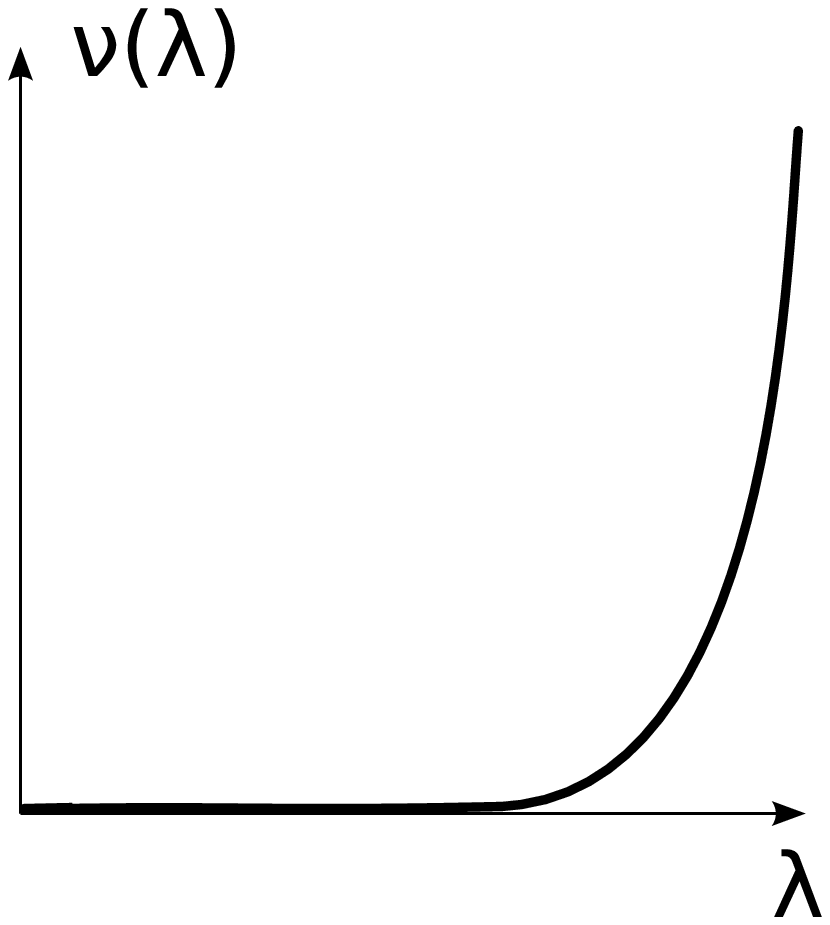}}
\caption{\label{spectrum} Fermionic spectrum of the chirally symmetric Dirac operator in a finite volume for $T<T_c$ (left), $T_c < T\lesssim 2\,T_c$ (center) and $T > 2\,T_c$ (right).}
\end{figure}

So far we are not able to describe precisely the microscopic structure of the collective field $\theta$. However, the fact that it may consist of the zero or near-zero fermionic modes can be deduced from various recent studies. By considering probe quarks one can show that the fermionic spectrum at the intermediate temperatures in interest ($T_c < T \lesssim 2\,T_c$) consists of near-zero modes and the bulk of the spectrum, separated from the former by a gap (see \cite{kalaydzhyan} and refs. therein for early and quenched studies, as well as \cite{Sharma:2013nva, Alexandru:2014tca} for the most recent results with dynamical fermions). This situation is sketched in Fig.~\ref{spectrum}. A near-zero peak in the second figure was expected since long ago from the interacting instanton liquid model \cite{Schafer:1995pz}. A long-wavelength mode slightly above $T_c$ was also obtained in a different way in Ref.~\cite{Gao:2014rqa}. This gives us more hints supporting the two-component model.

Regarding the space- and time-dependent picture, one of the possibilities might be the long-distance propagation of light quarks along the low-dimensional extended structures, populating the QCD vacuum (see, e.g. \cite{kalaydzhyan, Zakharov:2012vv, Buividovich:2011cv} and refs. therein), if the structures exist beyond the probe quark limit. Another study \cite{Chernodub:2012, Chernodub:2014hla} shows that, in the Copenhagen (``spaghetti'') model of the QCD vacuum, the fermionic zero modes are localized on the chromodynamic flux tubes and propagate along them in a direction specified by their chirality. This may give us a clue of the collective behavior of the quark zero modes.

{\bf Acknowledgements.}
I would like to thank DESY Theory Group, where most of this work was done, and Valentin~I.~Zakharov, Henry Verschelde, Mikhail Isachenkov and Frash\"{e}r Loshaj for numerous discussions. This work was supported in part by the U.S. Department of Energy under Contract No. DE-FG-88ER40388.

\end{document}